\documentclass[%
 reprint, 
 superscriptaddress, 
 amsmath,amssymb,
 aps, 
 prb, 
 longbibliography, 
]{revtex4-2}

\usepackage{graphicx}
\usepackage{dcolumn}
\usepackage{bm}
\usepackage[utf8]{inputenc}
\usepackage{makecell}

\begin{document}

\preprint{APS/123-QED}

\title{
\textbf{Graphical Rules for Zero-energy Quantum-Interference Patterns in Bipartite Lattices}
}

\author{Xinhao Fan}
\affiliation{%
Department of Physics, Shanghai University, Shanghai 200444, China.
}%
\author{Shunbo Hu}%
 \email{shunbohu@shu.edu.cn}
\affiliation{%
Institute for Quantum Science and Technology, Department of Physics, Institute for the Conservation of Cultural Heritage, Shanghai University,Shanghai 200444, China.
}%

\author{Jingzhe Chen}
\email{jingzhe@shu.edu.cn}
\affiliation{%
Department of Physics, Shanghai University, Shanghai 200444, China.
}%

\date{\today}

\begin{abstract}
Quantum interference (QI) plays an important role in electron transport in molecules, where conductance is determined by the coherent superposition of different quantum-walk paths connecting the source and drain sites. In some systems, the pronounced even–odd conductance oscillation with the walk length can be observed; meanwhile destructive QI can lead to a dramatic suppression of conductance, resulting in a huge on/off ratio. In this paper, we extend the investigation from individual molecules to bipartite lattice systems, providing an abstract framework for exploring QI in a broad range of two-dimensional systems. We propose a simple graphical rule for predicting the QI pattern—the conductance map obtained by varying the drain site while keeping the source site fixed—in bipartite structures with arbitrary boundaries and holes. The analysis can be completed in three steps. First, a unit-cell-replicated bipartite subset is constructed by removing a subset of atoms from the input network, and its Green's function map is efficiently obtained from a local recurrence relation. Second, the removed atoms and local structural modifications are coupled back as local degrees of freedom, generating zero-mode vectors through the same Green's function map. Finally, the surviving zero-mode combinations are selected by an effective coupling matrix, whose projectors correct the reference map to yield the final QI pattern. This rule applies to finite flakes with arbitrary boundaries and holes and extends naturally to infinite lattices and vacancy defects.
\end{abstract}

\maketitle


\section{Introduction}
\label{sec:introduction}

Quantum interference (QI) offers a powerful means of controlling coherent electron transport at the nanoscale. In molecular junctions, destructive QI can suppress conductance by several orders of magnitude, providing a quantum-mechanical route to high on/off ratios in molecular-scale devices~\cite{NN_Chen2024}. However, the structural origin of a transmission node is often difficult to identify directly from the underlying molecular structure. For realistic molecular devices, interpreting these transmission nodes may require Green's function calculations, electronic-structure analysis, or local-current mapping, and their occurrence can depend sensitively on both molecular connectivity and contact geometry ~\cite{Mujica1994, NC_Solomon2010, PedersenPRB, Lambert2015}. This difficulty has motivated graphical and graph-theoretical rules that relate destructive QI in conjugated molecules to simple connectivity criteria~\cite{Markussen2010,Markussen2011,Tsuji2018}. Such rules are valuable because they provide structural insight without requiring a full transport calculation for each contact configuration. In a two-dimensional lattice, the corresponding problem is naturally spatial: one can fix a source site and vary the drain site across the flake to obtain a conductance map. Therefore, a graphical rule for such systems should predict the full QI pattern, rather than merely indicate the presence or absence of a transmission node for a single pair of contacts.

Here we focus on bipartite lattices which are popular in two-dimensional materials. Graphene is a prominent example, for which the nearest-neighbor hopping approximation provides a simple and widely used description of the low-energy electronic structure while preserving the underlying bipartite connectivity. Its honeycomb structure, sublattice degree of freedom, and low-energy electronic properties make it a concrete platform for studying coherent transport on a lattice~\cite{Wallace1947,CastroNeto2009,DasSarma2011}. QI effects have been explored extensively in graphene nanostructures, graphene quantum junctions, nanoporous graphene, and local-probe geometries, where contact position, sublattice structure, and edge geometry can strongly affect transmission nodes and antiresonances~\cite{PRL_Jauho,Valli2018,Valli2019,Valli2023,JACS_Jauho}. These studies show that lattice connectivity and contact geometry can be used as design principles for coherent transport in graphene-based systems. Therefore, we use graphene as a representative example to illustrate the graphical rule below, although the underlying rule is not specific to graphene and is generally applicable to bipartite lattice structures.

In bipartite lattices, the nearest-neighbor hopping approximation gives rise to a chiral, block-off-diagonal Hamiltonian. Consequently, chiral symmetry imposes particularly strong constraints on the Green's function at zero energy. When the hopping block is nonsingular, this symmetry forces the same-sublattice Green's function elements to vanish at zero energy, while the opposite-sublattice elements are determined by the inverse of hopping block. When the hopping block is singular, zero modes emerge, producing poles in the Green's function. Zigzag edges in graphene nanoribbons host near-zero-energy boundary states~\cite{Nakada1996,Fujita1996,Wakabayashi2010}, while in finite graphene nanoislands, the number and structure of zero modes depend sensitively on shape, edge termination, and sublattice imbalance~\cite{Potasz2010,FernandezRossier2007,Wang2008,Kuc2010}. We note that, in the context of this work, zero modes refer to localized zero-energy states of the finite flake and is not able to conduct carriers. This does not preclude zero-energy transport, since coupling to the contacts broadens nearby low-energy propagating states and can give them finite spectral weight at zero energy. Generally, the chiral block structure of bipartite hopping Hamiltonians enforces a symmetric pairing of nonzero energies about zero, while unpaired zero-energy states can be chosen to reside entirely on a single sublattice~\cite{Coulson1940, LonguetHiggins1950}. Their number is bounded from below by the sublattice imbalance and can be further increased by rank deficiency associated with local connectivity ~\cite{Sutherland1986,Inui1994}. Vacancies and short-range defects provide additional routes to zero modes and resonant low-energy states in graphene~\cite{Pereira2006,Pereira2008,Wehling2010}.A graphical rule for finite bipartite lattice flakes must therefore capture both the zero-energy propagation pattern in the absence of zero modes and its modification when zero modes emerge.

\begin{figure}[t]
\centering
\includegraphics[width=0.98\columnwidth]{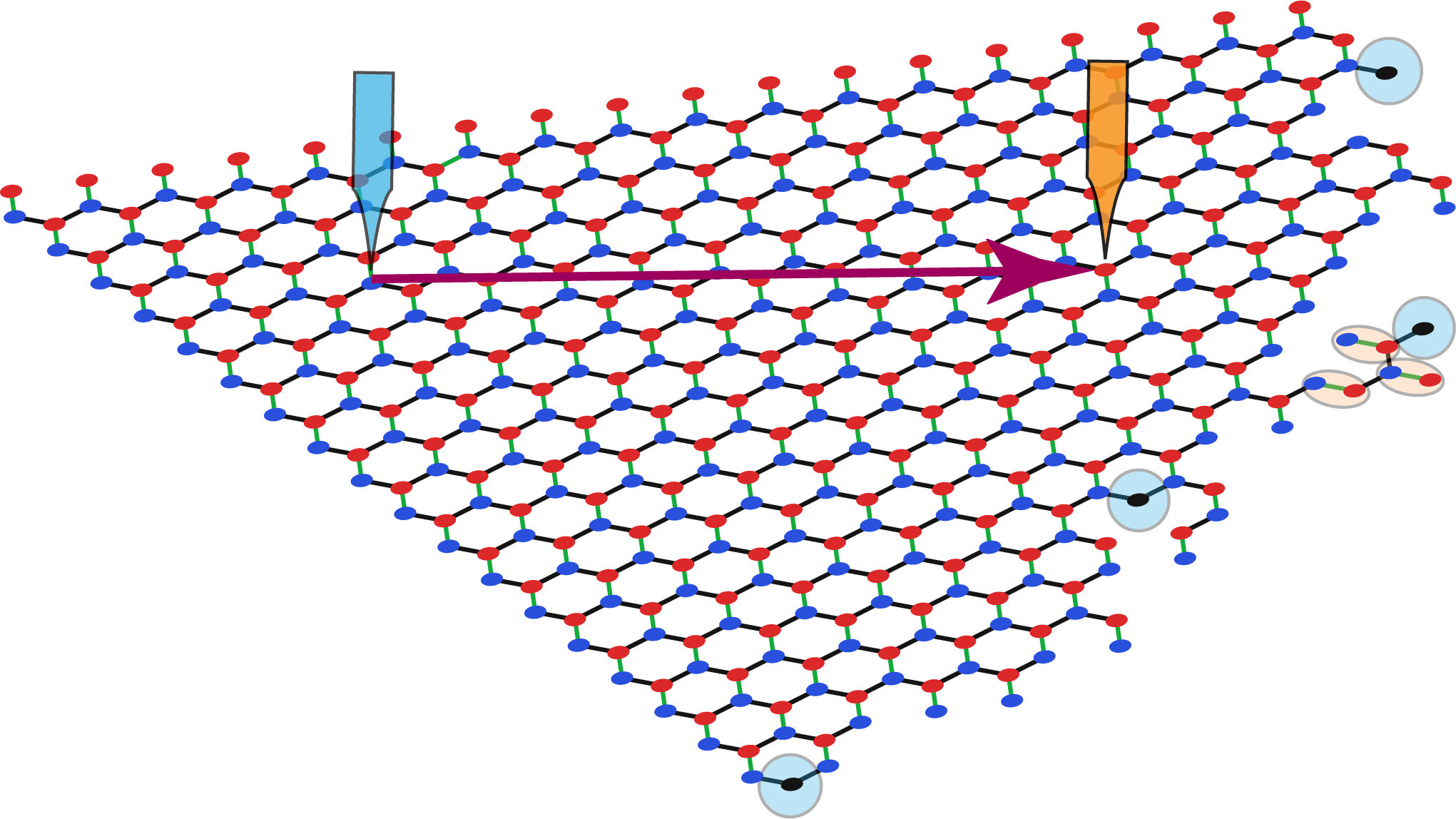}
\caption{Schematic of the finite flake and contacts considered in this work. Red and blue sites denote the A and B sublattices, respectively, in the unit-cell-replicated reference. Each red-blue pair connected by a green bond is assigned to a complete unit cell, and the collection of these cells defines the reference flake. The remaining black sites(shaded blue) represent local structural modifications, including attached atoms or incomplete boundary cells. The equivalent unit cells — which may differ visually from the original unit cell yet share identical hopping relations are shaded in orange. The two probes denote the source and drain electrodes, and the arrow indicates the direction of current flow. Green and black bonds represent the same nearest-neighbor hopping; green bonds are used solely to indicate the chosen unit-cell pairing.}
\label{fig:graphical_rule_schematic}
\end{figure}

In this work, we propose a graphical rule for constructing zero-energy QI maps in finite bipartite flakes. The analysis proceeds in three sequential steps. First, a unit-cell-replicated reference flake is constructed. Here, unit-cell replication means that the flake is generated from a single unit cell and its replicas without any other additional atoms. Fig.~\ref{fig:graphical_rule_schematic} illustrates the concept in an example of graphene, and we note that the regions shaded with transparent orange are equivalent to the original unit cell althrough the bond direction is different. In such a structure, one can easily get the isolated Green's function map through a simple recurrent relation. The term isolated Green's function map represents the propagation probability from the fixed source site to all the other possible drain sites in the flake without contacts. We note that it is
related to but different from the extended Green's function map with contacts which is used to calculate the final transmission. Second, the local structural modifications, which is the difference between the original flake and the unit-cell-replicated reference, are taken into account. Each atom in this subset itself leads to a zero-mode and opens/closes some channel in the previous isolate Green's function map. When there are more than one zero-modes the crossing interaction needs to be considered, some zero-modes can possibly be suppressed and the survival zero-modes give the complete zero-mode map. Finally, the isolate Green's function map of the original flake can be derived from the zero-mode map and renders the transmission through the exact two-contact identity. Through these three steps, one can achieve the transmission map purely from the structure. This analysis can also be extended to infinite lattices, with some change in technical details.

The paper is organized as follows. Section~\ref{sec:graphical_rule} introduces the bipartite tight-binding model and the effects of local probes, derives the zero-energy graphical recurrence for the unit-cell-replicated reference flake, and obtains the directed Yanghui-triangle(also named as Pascal-triangle) map. Section ~\ref{sec:local_modifications_zero_mode_correction} extends to general finite flakes by separating the local structural modifications, constructing the effective matrix $M$, generating the surviving zero-mode maps, and deriving the source-selected channel rules. Section~\ref{sec:infinite_lattice_local_defects} develops the corresponding reference map for infinite graphene and treats local defects through finite-rank corrections. Sections~\ref{sec:discussion_conclusion} and \ref{sec:conclusion} present the discussion and conclusions, respectively.

\section{Graphical rule for a unit-cell-replicated flake}
\label{sec:model_green-function_formulation}
\label{sec:graphical_rule}

We consider a finite bipartite flake with sublattices $A$ and $B$. In the nearest-neighbor tight-binding approximation, hopping occurs only between opposite sublattices, and the Hamiltonian takes the block-off-diagonal form
\begin{equation}
H_F=
\begin{pmatrix}
0 & h_F\\
h_F^\dagger & 0
\end{pmatrix},
\end{equation}
where $h_F:\mathcal H_B\rightarrow\mathcal H_A$ maps the single-particle orbital space of the $B$ sublattice to that of the $A$ sublattice. The retarded Green function of the isolated flake is
\begin{equation}
g_F(z)=(zI-H_F)^{-1},
\quad
z=E+i\eta,
\quad
\eta\rightarrow0^+,
\end{equation}
and its matrix elements are denoted by
\begin{equation}
g_{ij}(z)\equiv\langle i|g_F(z)|j\rangle.
\label{eq:isolated_matrix_elements}
\end{equation}

For a fixed source site $s$, the set of elements $g_{ds}$ obtained by scanning the drain site $d$ forms the isolated Green-function map that underlies the graphical construction.

To connect this map to transport, we attach local source and drain probes through
\begin{equation}
\Sigma_s(z)=\sigma_s(z)|s\rangle\langle s|,
\quad
\Sigma_d(z)=\sigma_d(z)|d\rangle\langle d|.
\end{equation}

The contacted Green function is
\begin{equation}
G_F(z)=\left[zI-H_F-\Sigma_s(z)-\Sigma_d(z)\right]^{-1},
\end{equation}
with matrix elements $G_{ij}(z)\equiv\langle i|G_F(z)|j\rangle$. For single-site contacts, the transmission is
\begin{align}
&T_{d\leftarrow s}(E)
=
\Gamma_s(E)\Gamma_d(E)|G_{ds}(E)|^2,
\notag\\
&\Gamma_{s,d}(E)
=-2\operatorname{Im}\sigma_{s,d}(E).
\label{eq:transmission_definition}
\end{align}
and Dyson's equation gives the exact two-contact identity
\begin{equation}
G_{ds}
=
\frac{g_{ds}}
{1-\sigma_s g_{ss}-\sigma_d g_{dd}
+\sigma_s\sigma_d\left(g_{ss}g_{dd}-g_{sd}g_{ds}\right)}.
\label{eq:two_contact_identity}
\end{equation}
Thus, the spatial transmission map can be constructed once the corresponding isolated Green-function elements are known~\cite{Caroli1971,FisherLee1981,Datta1995}.

We first consider a unit-cell-replicated flake constructed exclusively from complete bipartite unit cells, without individually added or removed sites or modified internal hoppings. Its hopping block is denoted by $h_0$. For the honeycomb connectivity considered here, $h_0$ is invertible; a concise proof is given in Appendix~\ref{app:h0_invertibility}. The reference Hamiltonian and its zero-energy Green function are therefore
\begin{equation}
H_0=
\begin{pmatrix}
0 & h_0\\
h_0^\dagger & 0
\end{pmatrix},
\quad
g_0(0)=
-\begin{pmatrix}
0 & {(h_0^{\dagger})}^{-1}\\
h_0^{-1} & 0
\end{pmatrix}.
\label{eq:zero_energy_inverse}
\end{equation}

Equation~\eqref{eq:zero_energy_inverse} leads to the first graphical rule: in the unit-cell-replicated reference at zero energy, once the source site is fixed on one sublattice, the isolated Green's function between the source and any site on the same sublattice vanishes identically, whereas that between the source and a site on the opposite sublattice is given by the corresponding matrix element of $h_0^{-1}$.

\begin{figure}[t]
\centering
\includegraphics[width=0.98\columnwidth]{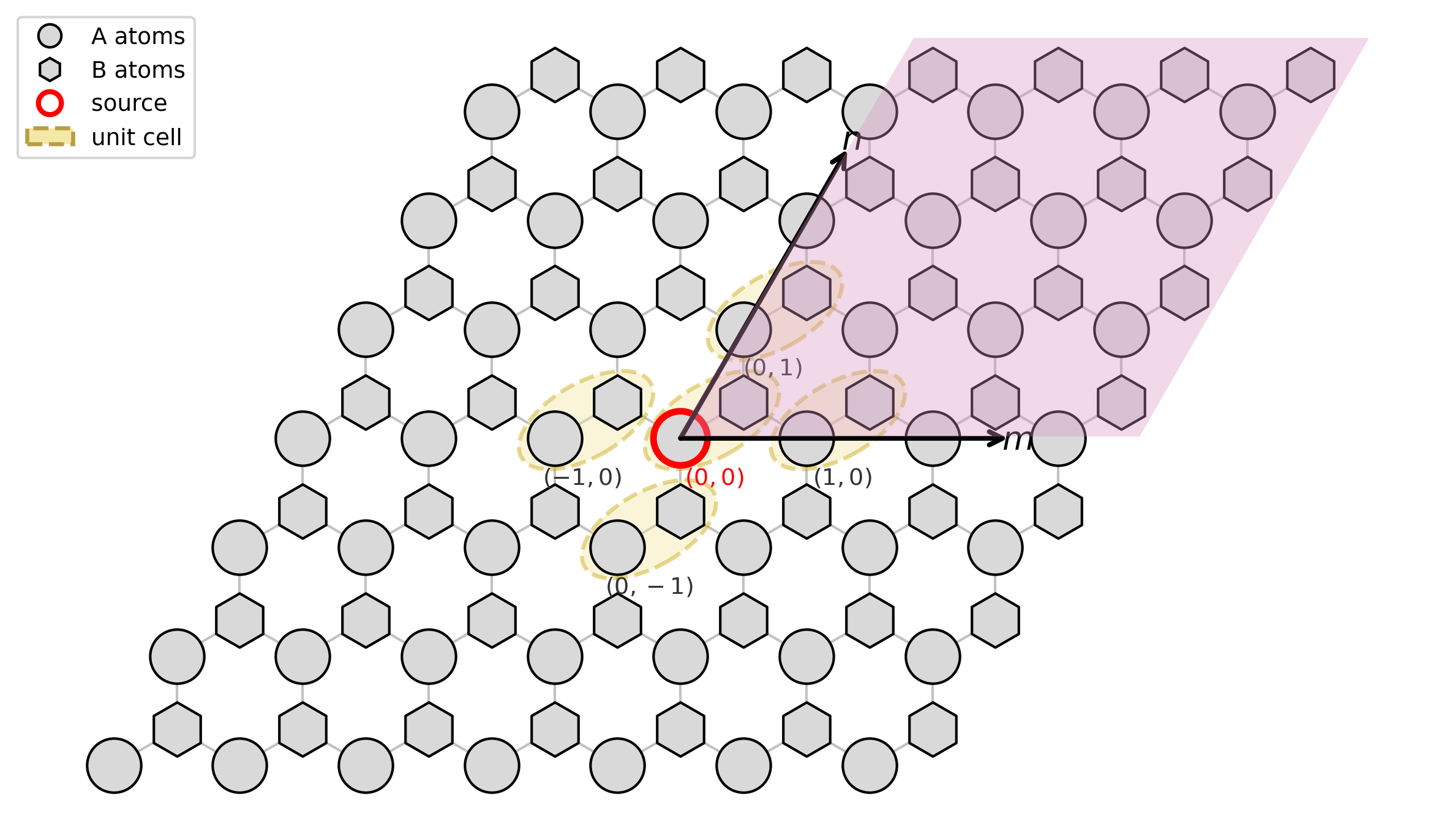}
\caption{
Coordinate convention for the graphene lattice.
Dashed ellipses indicate the chosen unit cells labeled by $(m,n)$, and the arrows mark the positive $m$ and $n$ directions. The shaded region is the directed sector.
}
\label{fig:graphene_coordinates}
\end{figure}

To construct these amplitudes without explicitly inverting $h_0$, let $a_{m,n}$ and $b_{m,n}$ denote the $A$ and $B$ sites of the unit cell indexed by $(m,n)$, and choose the source as $s=a_{0,0}$, as illustrated in Fig.~\ref{fig:graphene_coordinates}.

We define the source-selected Green-function column
\begin{equation}
|\mathcal A^{(0)}\rangle
\equiv
g_0(0)|a_{0,0}\rangle
=
\begin{pmatrix}
0\\
|\mathcal A_B^{(0)}\rangle
\end{pmatrix},
\end{equation}
where
\begin{align}
&|\mathcal A_B^{(0)}\rangle
=
\sum_{m,n}\mathcal A^{(0)}_{m,n}|b_{m,n}\rangle,
\notag\\
&\mathcal A^{(0)}_{m,n}
\equiv
\langle b_{m,n}|g_0(0)|a_{0,0}\rangle.
\label{eq:reference_amplitude_definition}
\end{align}
The equation $-H_0|\mathcal A^{(0)}\rangle=|a_{0,0}\rangle$ reduces to
\begin{equation}
-h_0|\mathcal A_B^{(0)}\rangle=|a_{0,0}\rangle.
\label{eq:reference_column_equation}
\end{equation}
With hopping amplitude $-t$, each $A$ site $a_{m,n}$ is connected to $b_{m,n}$, $b_{m-1,n}$, and $b_{m,n-1}$, and Eq.~\eqref{eq:reference_column_equation} becomes
\begin{equation}
t\left(
\mathcal A^{(0)}_{m,n}
+\mathcal A^{(0)}_{m-1,n}
+\mathcal A^{(0)}_{m,n-1}
\right)
=
\delta_{m0}\delta_{n0}.
\label{eq:local_recursion_minimal}
\end{equation}
If one of the neighboring sites is absent from the finite flake, its amplitude in Eq.~\eqref{eq:local_recursion_minimal} is taken as zero.

Equation~\eqref{eq:local_recursion_minimal} can be applied directly on the lattice. A non-source $A$ at the boundary site with only one neighbor forces the amplitude on that neighboring $B$ site to vanish. Repeating the same local sum rule along the adjacent boundary propagates these zeros along the zigzag edge and then inward. The source is the only site carrying an inhomogeneous term, so non-zero amplitudes are generated only after the inward recurrence reaches the source. We refer to the region subsequently reached by this source-generated recurrence as the directed sector(the shaded region in Fig.~\ref{fig:graphene_coordinates}).

Inside the directed sector, the recurrence has a closed form. Introducing the commuting shifts $L_m$ and $L_n$ through
\begin{equation}
h_0=-t(I+L_m+L_n),
\end{equation}
with
\begin{align}
&\langle b_{m,n}|L_m|\mathcal A_B^{(0)}\rangle
=\mathcal A^{(0)}_{m-1,n},
\notag\\
&\langle b_{m,n}|L_n|\mathcal A_B^{(0)}\rangle
=\mathcal A^{(0)}_{m,n-1}.
\end{align}
we note that the finite boundary makes the shifts nilpotent. Hence
\begin{equation}
h_0^{-1}
=
-\frac{1}{t}
\sum_{\ell=0}^{p-1}
\left[-(L_m+L_n)\right]^\ell,
\end{equation}
for some positive integer $p$. The directionality is therefore determined by the finite boundary. This local construction is closely related to Green's-function descriptions and path-sum descriptions of quantum walks on graphs~\cite{Andrade2011PRA,Andrade2016}. Within an untruncated part of the directed sector, the shifts commute along the contributing paths, and acting on the source gives the Yanghui-triangle rule
\begin{equation}
\mathcal A^{(0)}_{m,n}
=
\frac{(-1)^{m+n}}{t}
\binom{m+n}{m}
\label{eq:pascal_rule}
\end{equation}
within the directed sector.

\begin{figure}[t]
\centering
\includegraphics[width=0.98\columnwidth]{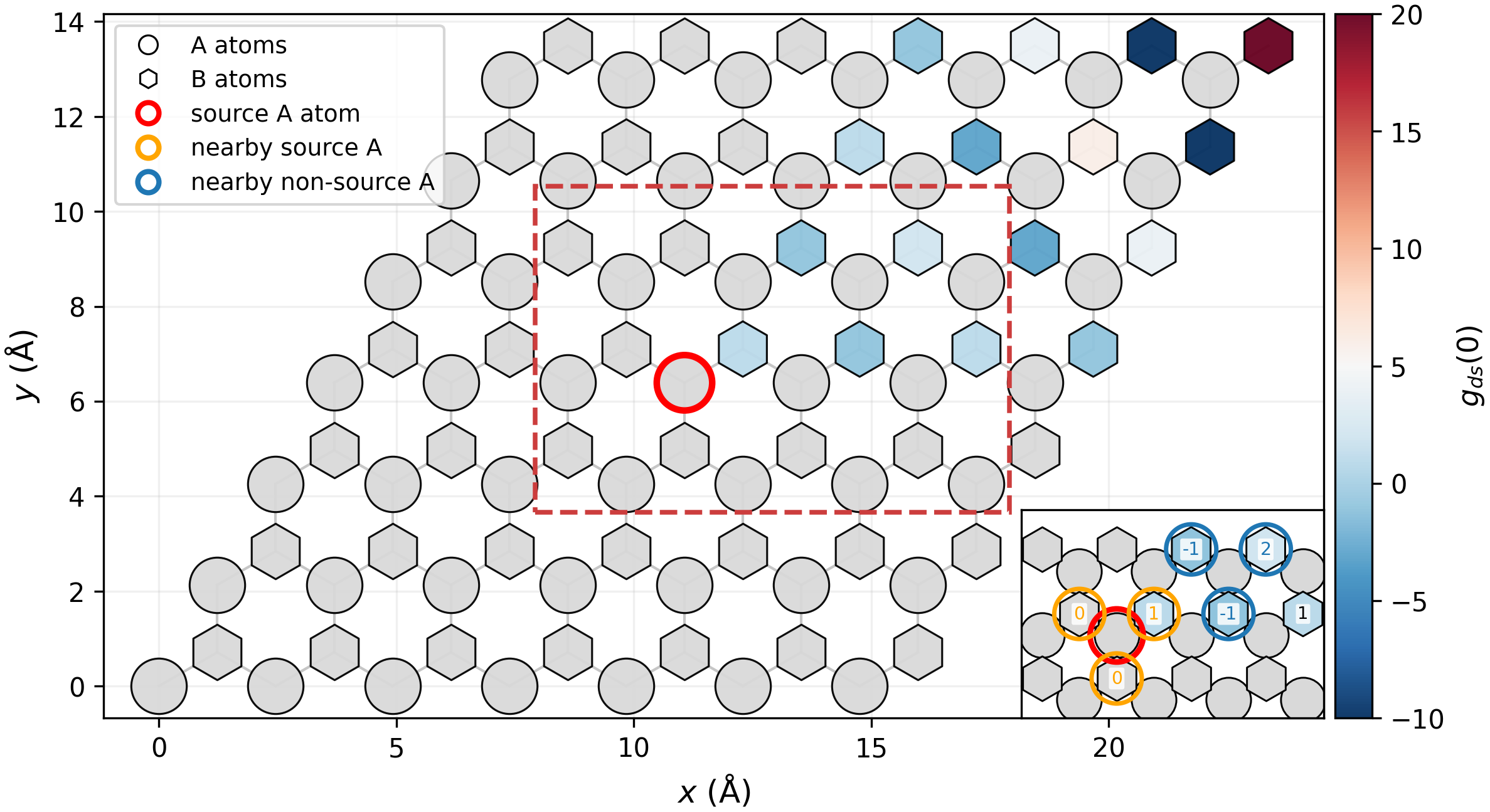}
\caption{Zero-energy reference Green-function map for a regular-shaped unit-cell-replicated flake. The source-centered amplitudes in the directed sector follow the Yanghui-triangle rule of Eq.~\eqref{eq:pascal_rule}.}
\label{fig:regular_gds_map}
\end{figure}

Figure~\ref{fig:regular_gds_map} shows the resulting map for a regular-shaped unit-cell-replicated flake. Starting from the boundary-imposed zeros, the local recurrence generates the alternating binomial amplitudes of Eq.~\eqref{eq:pascal_rule} throughout the directed sector.

\begin{figure}[t]
\centering
\includegraphics[width=0.98\columnwidth]{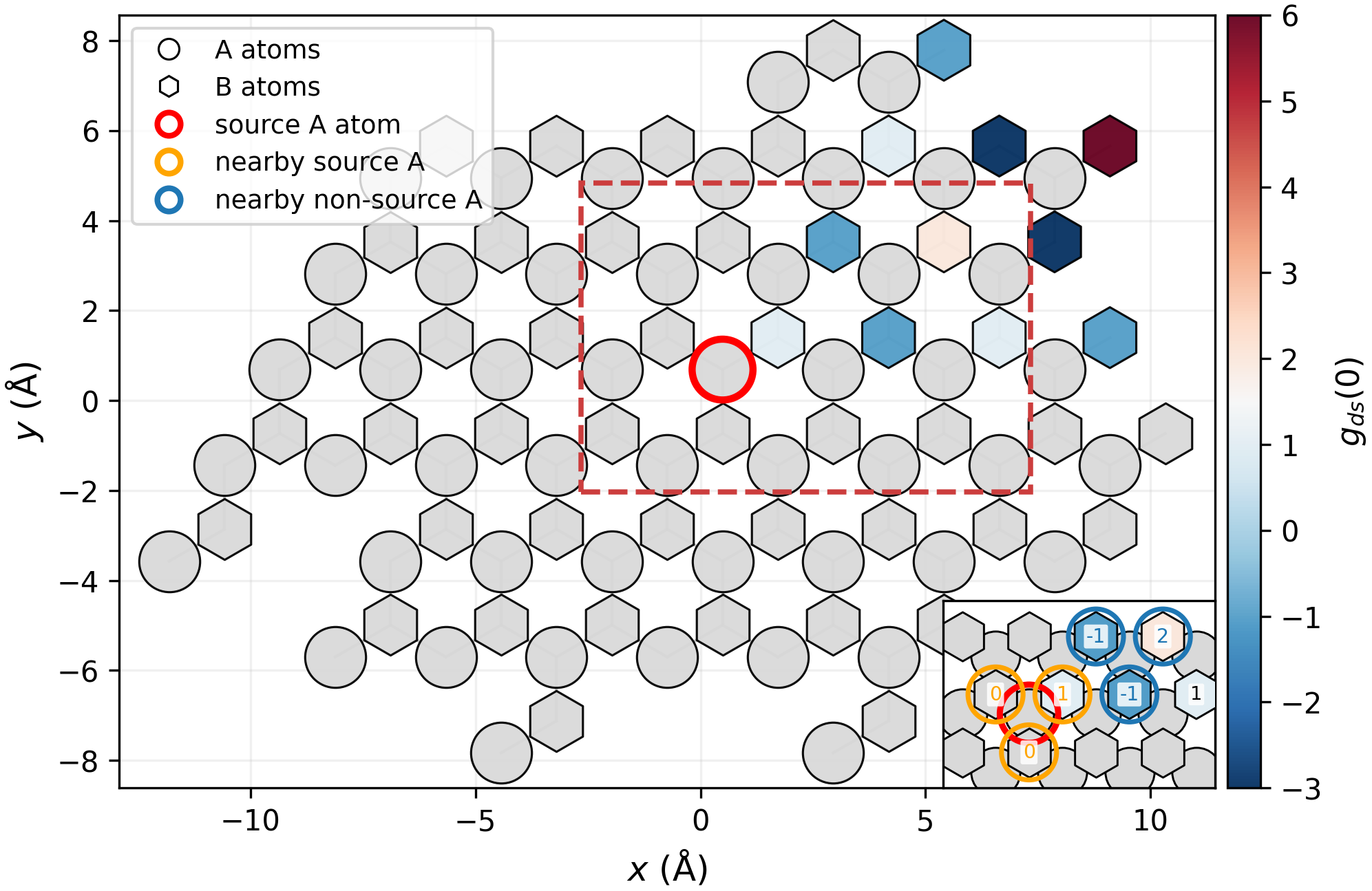}
\caption{Zero-energy reference Green-function map for a random-shaped unit-cell-replicated flake. The irregular boundary truncates the accessible region but leaves the locally generated amplitudes unchanged wherever the directed sector overlaps that of the regular-shaped flake.}
\label{fig:random_gds_map}
\end{figure}

The graphical construction does not require a globally regular outer boundary. The same recurrence can be applied to any flake assembled entirely from complete unit cells. The Yanghui-triangle distribution is therefore shape independent within their common untruncated directed sector. Figure~\ref{fig:random_gds_map} illustrates this point for a random-shaped boundary. 

The reference Green-function map is substituded into a transmission map through Eq.~\eqref{eq:two_contact_identity}. With the symmetric wide-band limit, $\sigma_s=\sigma_d=-i|t|$, the diagonal isolated Green-function elements vanish and Eq.~\eqref{eq:pascal_rule} gives

\begin{equation}
T_{b_{m,n}\leftarrow a_{0,0}}(0)
=
\frac{4\left|\binom{m+n}{m}\right|^2}
{\left[1+\left|\binom{m+n}{m}\right|^2\right]^2}.
\label{eq:reference_transmission_rule}
\end{equation}

Thus, the reference construction is completed. A general flake requires one additional step: the sites out of the unit-cell-replicated structure must be reintroduced as local degrees of freedom.

\section{Graphical rule for general finite flakes}
\label{sec:local_modifications_zero_mode_correction}

Section~\ref{sec:graphical_rule} provides the source-driven Green-function map for a unit-cell-replicated flake. To extend this construction to a general finite flake, we retain the maximal unit-cell-replicated subset as the reference and treat all remaining sites as local degrees of freedom. The graphical procedure then has two steps: the local structure is first used to construct the zero-mode maps, after which the source position selects the same- or opposite-sublattice branch that contributes to the final transmission map.

\subsection{Local modifications and zero-mode construction}

Let $A_0$ and $B_0$ denote the two sublattices of the reference flake, and let $A_r$ and $B_r$ contain the remaining local degrees of freedom. The reference set needs not be geometrically connected; the essential requirement is that its hopping block $h_0$ is invertible. We denote the local basis states by $\{|A_{r,\ell}\rangle\}_{\ell=1}^{p}$ and $\{|B_{r,j}\rangle\}_{j=1}^{q}$, where $p=\dim\mathcal H_{A_r}$ and $q=\dim\mathcal H_{B_r}$. In the ordered bases $(A_0,A_r)$ and $(B_0,B_r)$, the full hopping block is
\begin{equation}
h_F=
\begin{pmatrix}
h_0 & W_B\\
W_A & X
\end{pmatrix},
\label{eq:full_hopping_block_local}
\end{equation}
where $W_A$ is the hopping block between $A_r$ and $B_0$ and $W_B$ is the hopping block between $A_0$ and $B_r$. It is convenient to write
\begin{equation}
W_A=\sum_{\ell=1}^{p}|A_{r,\ell}\rangle\langle w_\ell^A|,
\quad
W_B=\sum_{j=1}^{q}|w_j^B\rangle\langle B_{r,j}|,
\label{eq:local_coupling_vectors}
\end{equation}
where $|w_l^A\rangle$ and $|w_j^B\rangle$ are the corresponding row and column for a single local site, respectively. For example, if the local site is connected to a single site,  we have $\langle w_\ell^A|=\tau_{a\ell}\langle b_\ell|$ and $|w_j^B\rangle=\tau_{bj}|a_j\rangle$. And for the case of two connected sites, they are coherent sums as the following,
\begin{align}
\langle w_\ell^A|=\tau_{a\ell_1}\langle b_{\ell_1}|+\tau_{a\ell_2}\langle b_{\ell_2}|,
\notag\\
|w_j^B\rangle=\tau_{bj_1}|a_{j_1}\rangle+\tau_{bj_2}|a_{j_2}\rangle.
\label{eq:two_point_local_couplings}
\end{align} 
$X$ in Eq.~\ref{eq:full_hopping_block_local} is the hopping block between the local $A_r$ and $B_r$ sites. 

The local structure is encoded in the local effective coupling matrix
\begin{equation}
M\equiv S-X,
\quad
S\equiv W_Ah_0^{-1}W_B,
\label{eq:local_effective_matrices}
\end{equation}
where the submatrix $S$ represents the coupling effect from the reference flake with its element written as
\begin{equation}
S_{\ell j}=\tau_{a\ell}\tau_{bj}\langle b_\ell|h_0^{-1}|a_j\rangle.
\label{eq:S_matrix_element}
\end{equation}
The null spaces of $M^{\dagger}$ and $M$ determine the surviving $A$- and $B$-sublattice zero modes, respectively,
\begin{equation}
M^\dagger|\alpha\rangle=0,
\quad
M|\beta\rangle=0.
\label{eq:zero_mode_selection_conditions}
\end{equation}
The components of $|\alpha\rangle$ and $|\beta\rangle$ are the relative zero-mode amplitudes on the local $A_r$ and $B_r$ sites respectively. The corresponding zero states can be solved as
\begin{align}
|\varphi^A\rangle&=
\begin{pmatrix}
-h_0^{-\dagger}W_A^\dagger|\alpha\rangle\\
|\alpha\rangle
\end{pmatrix},
\notag\\
|\varphi^B\rangle&=
\begin{pmatrix}
-h_0^{-1}W_B|\beta\rangle\\
|\beta\rangle
\end{pmatrix}.
\label{eq:surviving_zero_modes}
\end{align}
If $r=\operatorname{rank}M$, the numbers of zero modes are
\begin{equation}
n_0^A=p-r,
\quad
n_0^B=q-r.
\label{eq:number_zero_modes}
\end{equation}

For an $A$-sublattice zero mode, the zero-energy equation $H_F|\varphi^A\rangle=0$ can be solved in a similar recursive way as that in Sec.~\ref{sec:graphical_rule}. The corresponding $A$- and $B$-sublattice recurrences are

\begin{align}
&t\left(
\varphi^A_{m,n}
+\varphi^A_{m+1,n}
+\varphi^A_{m,n+1}
\right)=0,
\notag\\
&t\left(
\varphi^B_{m,n}
+\varphi^B_{m-1,n}
+\varphi^B_{m,n-1}
\right)=0.
\label{eq:homogeneous_zero_mode_recurrence}
\end{align}
respectively with the same unit-cell indexing definition.

\begin{figure}[t]
\centering
\includegraphics[width=0.98\columnwidth]{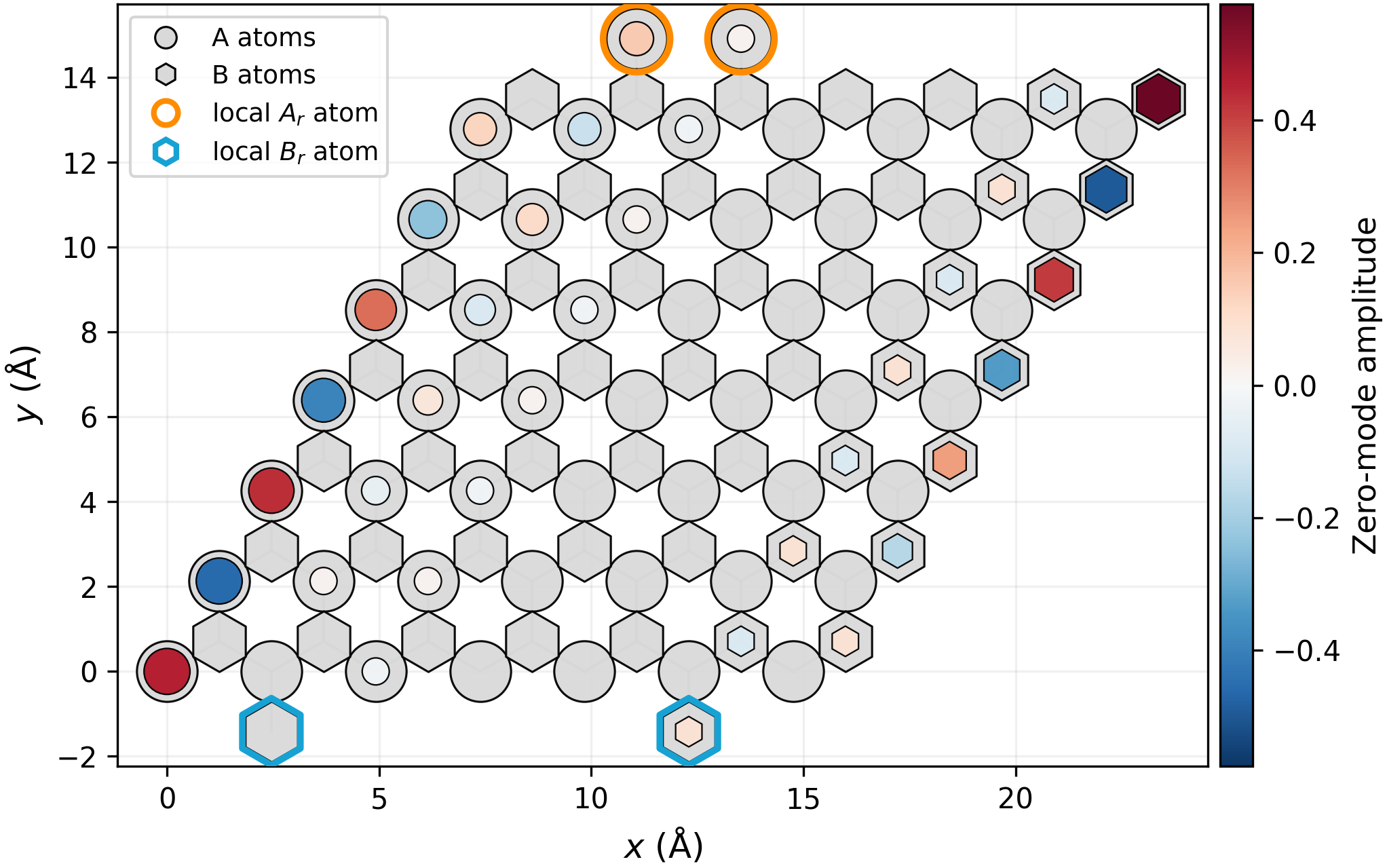}
\caption{Normalized signed amplitudes of the single surviving $A$- and $B$-sublattice zero modes of the locally modified graphene flake. Circles and hexagons show the $A$- and $B$-sublattice zero-mode amplitudes, respectively. Orange circles and blue hexagons identify the two local $A_r$ and two local $B_r$ sites.}
\label{fig:local_zero_modes}
\end{figure}

As an example, we consider the structure shown in Figure~\ref{fig:local_zero_modes}. The flake contains two local $A_r$ sites and two local $B_r$ sites ($p=q=2$). There is no direct hopping between these sites, giving $X=0$ and $M=S$. We label the local sites on each sublattice from left to right. Let $a_j$ denote the reference $A$ site connected to $B_{r,j}$, and $b_\ell$ the reference $B$ site connected to $A_{r,\ell}$. According to Eq.~\eqref{eq:S_matrix_element}, column $j$ of $S$ is determined by the reference propagation amplitudes from $a_j$ to $b_1$ and $b_2$. Both upper connection sites lie within the directed sector generated from the lower-left connection site $a_1$, where the recurrence of Sec.~\ref{sec:graphical_rule} gives nonzero amplitudes. In contrast, both lie outside the directed sector generated from the lower-right connection site $a_2$, so the corresponding reference amplitudes vanish. Consequently, only the first column of $S$ is nonzero:
\begin{equation}
S=
\begin{pmatrix}
s_{11} & 0\\
s_{21} & 0
\end{pmatrix},
\quad
s_{11},s_{21}\neq0.
\label{eq:example_S_structure}
\end{equation}
The zero--nonzero pattern alone gives $\operatorname{rank}M=1$ and hence
\begin{equation}
n_0^A=1,
\quad
n_0^B=1.
\label{eq:example_zero_mode_count}
\end{equation}
The corresponding local zero-mode amplitudes satisfy
\begin{equation}
|\alpha\rangle\propto
\begin{pmatrix}s^*_{21}\\-s^*_{11}\end{pmatrix},
\quad
|\beta\rangle\propto
\begin{pmatrix}0\\1\end{pmatrix}.
\label{eq:example_local_zero_mode_amplitudes}
\end{equation}
Thus, the zero--nonzero structure is sufficient to count the modes, while the ratio $s^*_{11}/s^*_{21}$ fixes the relative amplitudes on the two local $A_r$ sites. Propagating these local values with Eq.~\eqref{eq:homogeneous_zero_mode_recurrence} gives the normalized zero-mode profiles shown in Fig.~\ref{fig:local_zero_modes}.

\subsection{Source-selected channel maps}

Once the zero-mode maps have been constructed, the source position determines which part of the zero-energy Green function contributes to transport. Collect the zero modes into space $\Phi_{A/B}=(|\varphi_1^{A/B}\rangle,\ldots,|\varphi_{n_0^{A/B}}^{A/B}\rangle)$ and the propagator is 
\begin{equation}
\Pi_0^{A/B}=\Phi_{A/B}(\Phi_{A/B}^\dagger\Phi_{A/B})^{-1}\Phi_{A/B}^\dagger,
\label{eq:zero_mode_projectors}
\end{equation}
with $\Pi_0^{A/B}=0$ when no zero mode exists on the corresponding sublattice. The zero-mode spaces of A and B sublattices are independant of each other which can be easily proved from the eigen-equation, therefore near zero energy, the isolate Green's function is 
\begin{equation}
g_F(z)=\frac{\Pi_0}{z}+g_F^{\rm reg}(0)+O(z),
\quad
\Pi_0=\Pi_0^A+\Pi_0^B.
\label{eq:zero_mode_green_decomposition}
\end{equation}

\begin{table*}[t]
\caption{Zero-energy channel rules for a source $s_A$ on the $A$ sublattice. ``Open'' denotes a zero-mode-mediated channel, whereas ``Allowed'' means that the channel is not forbidden by the zero-mode structure; its amplitude is generated by the source-driven recurrence and may still vanish at an ordinary interference node.}
\label{tab:general_A_source_channel_rule}
\centering
\begin{ruledtabular}
\begin{tabular}{cccc}
\makecell[c]{Source\\condition}
&
Drain
&
\makecell[c]{Additional drain\\condition}
&
\makecell[c]{Zero-energy\\channel}
\\
\hline
$\mathbf p_s^A\neq0$
&
$d_A$
&
\makecell[c]{$\mathbf p_d^A\neq0$ and\\$\mathbf p_d^A\parallel\mathbf p_s^A$}
&
\makecell[c]{Open through the\\$A$-sublattice zero modes}
\\
$\mathbf p_s^A\neq0$
&
$d_A$
&
Otherwise
&
Closed
\\
$\mathbf p_s^A\neq0$
&
$d_B$
&
Any $d_B$
&
Closed
\\
$\mathbf p_s^A=0$
&
$d_A$
&
Any $d_A$
&
Closed
\\
$\mathbf p_s^A=0$
&
$d_B$
&
$\mathbf p_d^B=0$
&
\makecell[c]{Allowed; obtained from the\\source-driven recurrence}
\\
$\mathbf p_s^A=0$
&
$d_B$
&
$\mathbf p_d^B\neq0$
&
Closed
\end{tabular}
\end{ruledtabular}
\end{table*}

\begin{figure*}[t]
\centering
\begin{minipage}[t]{0.48\textwidth}
\centering
\includegraphics[width=\linewidth]{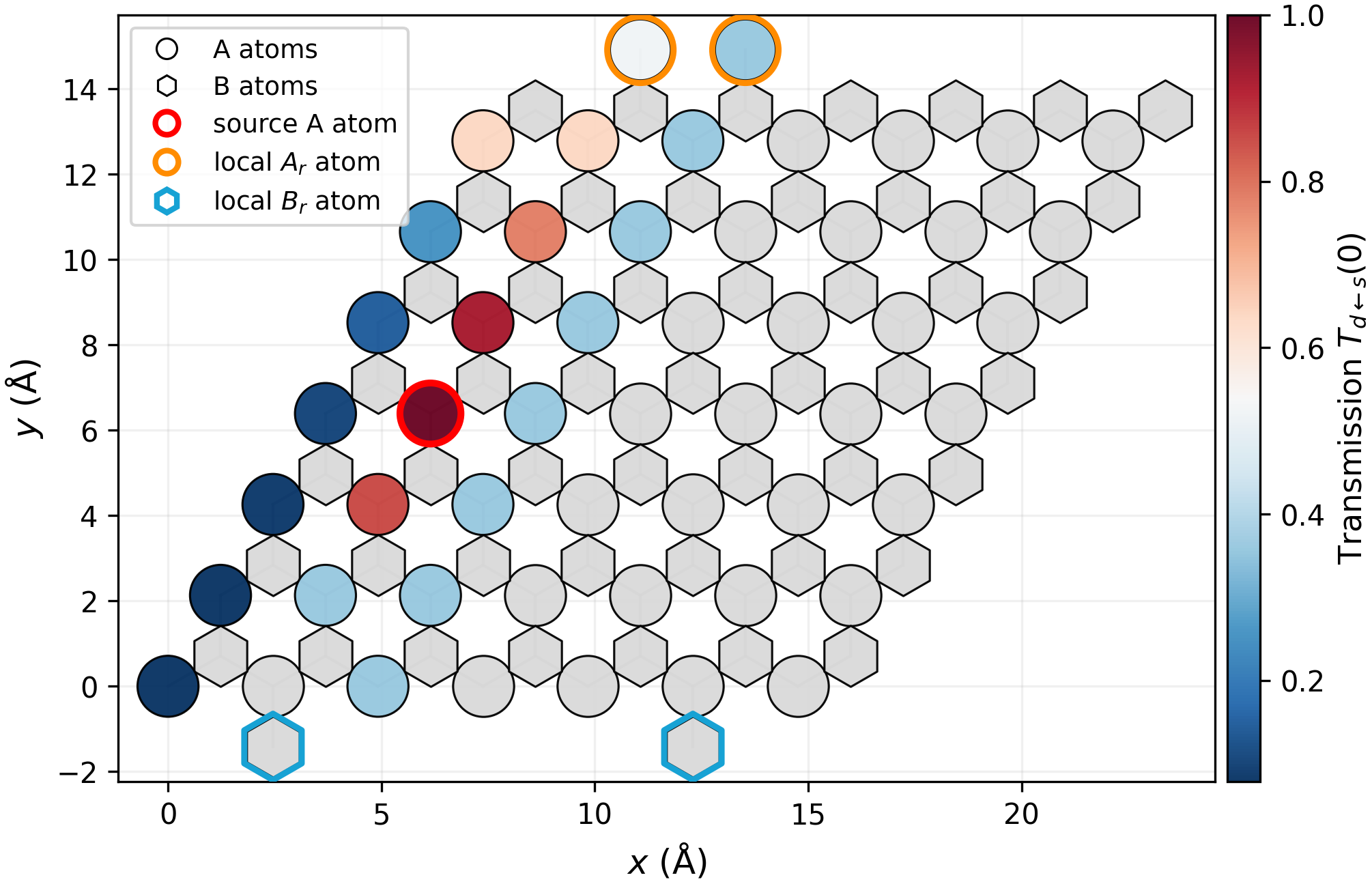}
\par{\small (a) The source lies within the support of the $A$-sublattice zero mode}
\end{minipage}
\hfill
\begin{minipage}[t]{0.48\textwidth}
\centering
\includegraphics[width=\linewidth]{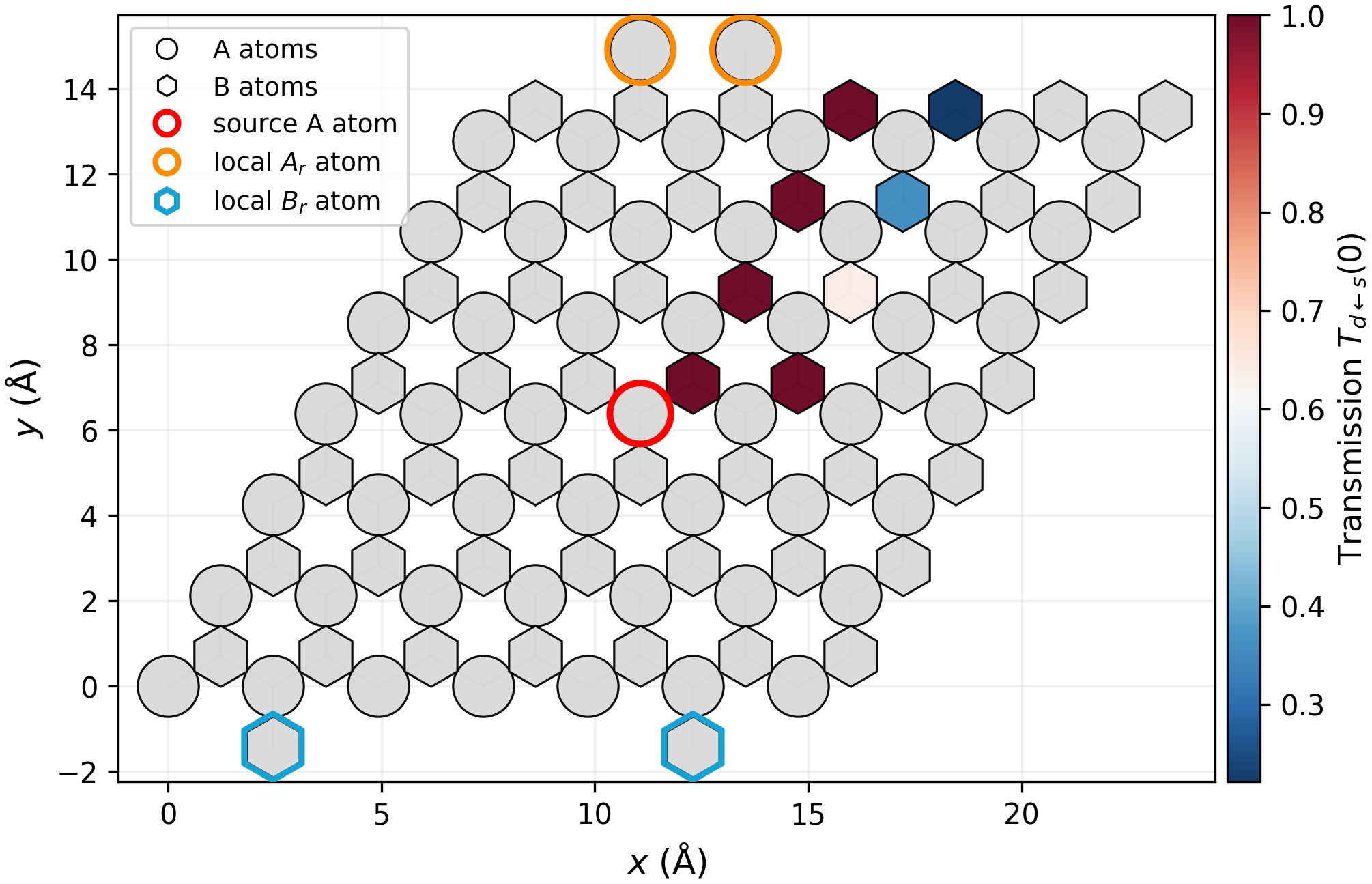}
\par{\small (b) The source lies outside the $A$-sublattice zero-mode support}
\end{minipage}
\caption{Zero-energy transmission maps for two $A$-sublattice source positions. (a) The source overlaps the $A$-sublattice zero mode, so the map is confined to the zero-mode-mediated same-sublattice channels and every opposite-sublattice channel is closed. (b) The source lies outside the $A$-zero-mode support, so the same-sublattice map is closed and the opposite-sublattice amplitudes are generated by the source-driven recurrence and retained only outside the $B$-zero-mode support.}
\label{fig:source_position_comparison}
\end{figure*}

Then we project the zero-mode states onto the sites connectted to local modifications, 
\begin{equation}
\mathbf p_x^A\equiv\Phi_A^\dagger|x_A\rangle,
\quad
\mathbf p_x^B\equiv\Phi_B^\dagger|x_B\rangle.
\label{eq:contact_zero_mode_projections}
\end{equation}
For an $A$-sublattice source, the analysis of the two-contact identity gives two branches. A same-sublattice channel can remain open only when
\begin{equation}
\mathbf p_s^A\neq0,
\quad
\mathbf p_d^A\neq0,
\quad
\mathbf p_d^A=c\,\mathbf p_s^A
\quad(c\neq0),
\label{eq:same_sublattice_channel_rule}
\end{equation}
in which case the relevant isolated Green-function pole is
\begin{equation}
g_{d_As_A}(z)=\frac{\langle d_A|\Pi_0^A|s_A\rangle}{z}+O(1).
\label{eq:same_sublattice_projector_element}
\end{equation}
If instead $\mathbf p_s^A=0$, every same-sublattice channel is closed and the opposite-sublattice branch is allowed only when
\begin{equation}
\Pi_0^A|s_A\rangle=0,
\quad
\Pi_0^B|d_B\rangle=0.
\label{eq:opposite_sublattice_channel_condition}
\end{equation}
For this branch, the source term is unchanged and the amplitudes are generated by the same source-driven local recurrence as in Sec.~\ref{sec:graphical_rule}, with the local connection conditions supplied by $W_A$, $W_B$, and $X$. The zero modes on the drain sublattice identify the sites that must be excluded. The derivation of these conditions and the full projected regular solution are given in Appendix~\ref{app:zero_mode_channels}.

The rules are summarized in Table~\ref{tab:general_A_source_channel_rule}. The result for a $B$-sublattice source follows by interchanging $A$ and $B$. The example in Fig.~\ref{fig:local_zero_modes} contains one normalized zero mode on each sublattice, denoted by $|\varphi^A\rangle$ and $|\varphi^B\rangle$. The projection vectors are then scalars, so the collinearity condition in Eq.~\eqref{eq:same_sublattice_channel_rule} holds whenever both zero-mode amplitudes are nonzero. Thus the graphical rule reduces to Table~\ref{tab:single_zero_mode_A_source}.

\begin{table}[t]
\caption{Zero-energy channel rules for an $A$-sublattice source when the full flake supports one zero mode on each sublattice. ``Allowed'' means that the channel is not excluded by the zero-mode structure; additional nodes may still arise from the source-driven recurrence.}
\label{tab:single_zero_mode_A_source}
\centering
\begin{ruledtabular}
\begin{tabular}{ccc}
\makecell[c]{Source\\position}
&
\makecell[c]{$A$-sublattice\\drains}
&
\makecell[c]{$B$-sublattice\\drains}
\\
\hline
$\phi^A(s_A)\neq0$
&
\makecell[c]{Open where\\$\phi^A(d_A)\neq0$}
&
\makecell[c]{Closed\\everywhere}
\\
$\phi^A(s_A)=0$
&
\makecell[c]{Closed\\everywhere}
&
\makecell[c]{Allowed only where\\$\phi^B(d_B)=0$}
\end{tabular}
\end{ruledtabular}
\end{table}

The following two source positions therefore activate different graphical branches. If the source lies within the scope of the $A$-sublattice zero mode, the same-sublattice map is read from the zero-mode projector and every opposite-sublattice channel is closed. Otherwise the same-sublattice map is closed; the opposite-sublattice amplitudes are generated by the ordinary source-driven recurrence and retained only at drains outside the scope of the $B$-sublattice zero mode. Figure~\ref{fig:source_position_comparison} shows the two cases for the same flake. The required isolated Green-function elements are finally inserted into Eq.~\eqref{eq:two_contact_identity} to obtain the transmission maps.

\section{Infinite-lattice reference map and local defects}
\label{sec:infinite_lattice_local_defects}

\subsection{Pristine infinite graphene}

The finite construction is organized around a reference Green-function map. In a finite flake, the boundary terminates the local recurrence and selects the directed sector. In an infinite lattice, the boundary is absent, so the corresponding reference map is instead fixed by translational symmetry.

For the infinite honeycomb lattice, let $\mathbf a_1$ and $\mathbf a_2$ be the Bravais primitive vectors and define
\begin{equation}
k_1\equiv\mathbf k\cdot\mathbf a_1,
\quad
k_2\equiv\mathbf k\cdot\mathbf a_2.
\end{equation}
In the sublattice basis, the Bloch Hamiltonian is
\begin{equation}
H_0^\infty(\mathbf k)
=
\begin{pmatrix}
0 & h(\mathbf k)\\
h^*(\mathbf k) & 0
\end{pmatrix},
\label{eq:infinite_bloch_hamiltonian}
\end{equation}
with
\begin{equation}
h(\mathbf k)
=-t\left(1+e^{-ik_1}+e^{-ik_2}\right).
\end{equation}
The momentum-space Green function is
\begin{equation}
g_0^\infty(\mathbf k,z)
=
\frac{1}{z^2-|h(\mathbf k)|^2}
\begin{pmatrix}
z & h(\mathbf k)\\
h^*(\mathbf k) & z
\end{pmatrix}.
\end{equation}
For a source at $a_{0,0}$ and a target at $b_{m,n}$,
\begin{equation}
g_{0}^{\infty;BA}(m,n;z)
=
\int_{\rm BZ}
\frac{dk_1dk_2}{4\pi^2}
\frac{e^{i(k_1m+k_2n)}h^*(\mathbf k)}
{z^2-|h(\mathbf k)|^2},
\label{eq:infinite_BA_green}
\end{equation}
where BZ denotes the first Brillouin zone. Its retarded zero-energy limit is
\begin{align}
\mathcal G(m,n)
&\equiv
g_{0}^{\infty;BA}(m,n;0)
\notag\\
&=
\frac{1}{t}
\int_{\rm BZ}
\frac{dk_1dk_2}{4\pi^2}
\frac{e^{i(k_1m+k_2n)}}
{1+e^{-ik_1}+e^{-ik_2}}.
\label{eq:infinite_reference_map_definition}
\end{align}
with $z=i\eta$ taken to zero only after the integration.

The lattice symmetry is conveniently expressed through
\begin{equation}
q_1=m,
\quad
q_2=n,
\quad
q_3=-m-n-1.
\end{equation}
A threefold lattice rotation $C_3$ cyclically permutes $(q_1,q_2,q_3)$, so that
\begin{equation}
\mathcal G(m,n)
=
\mathcal G(-m-n-1,m)
=
\mathcal G(n,-m-n-1).
\label{eq:C3_cyclic_green}
\end{equation}
The full lattice is covered by two inequivalent types of region: an acute sector, in which only one of the three $q_i$ is negative, and the region between two neighboring acute sectors, in which two are negative. Closed-form expressions for both regions are given in Appendix~\ref{app:infinite_closed_forms}. The same-sublattice blocks vanish in the retarded zero-energy limit, while the opposite-sublattice blocks obey
\begin{align}
&g_{0}^{\infty;AA}(\mathbf R;0)
=
g_{0}^{\infty;BB}(\mathbf R;0)
=0,
\notag\\
&g_{0}^{\infty;AB}(\mathbf R;0)
=
\left[g_{0}^{\infty;BA}(-\mathbf R;0)\right]^*.
\label{eq:infinite_sublattice_relations}
\end{align}
where $\mathbf R$ is the discrete lattice displacement from source to target.

\begin{figure}[t]
\centering
\includegraphics[width=0.98\columnwidth]{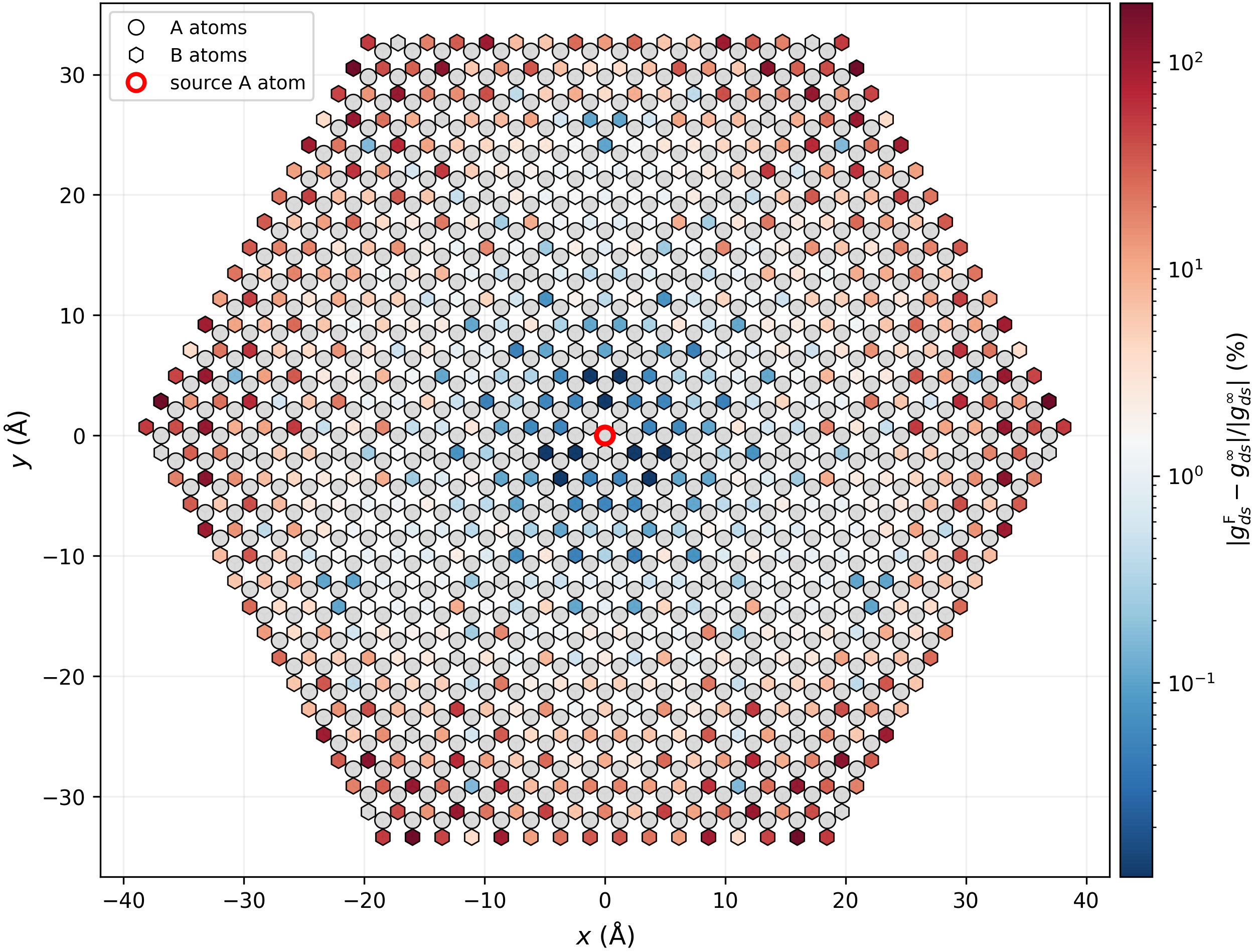}
\caption{
Relative error between the zero-energy isolated Green-function amplitudes of a finite $C_3$-symmetric graphene flake and pristine infinite graphene for a central $A$-sublattice source. Gray sites are excluded from the relative error evaluation because the corresponding infinite-lattice amplitude is below the adopted numerical threshold.
}
\label{fig:finite_infinite_relative_error}
\end{figure}

The infinite-lattice propagator plays the same role as the inverse hopping block of the finite reference flake. In particular,
\begin{equation}
\langle b|h_0^{-1}|a\rangle
\longrightarrow
-\,g_{0}^{\infty;BA}(\mathbf R_b-\mathbf R_a;0),
\label{eq:finite_to_infinite_reference_map}
\end{equation}
so the absence of a finite boundary changes how the reference map is obtained, but not how it is used. To compare the two reference constructions, we consider a finite $C_3$-symmetric graphene flake with an $A$-sublattice source at its center. We denote by $g_{ds}^{F}(0)$ the zero-energy isolated Green-function amplitude from the source $s$ to a drain $d$ in the finite flake, and by $g_{ds}^{\infty}(0)$ the corresponding amplitude in pristine infinite graphene for the same source--drain displacement. For each $B$-sublattice drain, their relative deviation is defined as
\begin{equation}
\delta_{ds}
=
\frac{\left|g_{ds}^{F}(0)-g_{ds}^{\infty}(0)\right|}
{\left|g_{ds}^{\infty}(0)\right|}.
\label{eq:finite_infinite_relative_error}
\end{equation}
Figure~\ref{fig:finite_infinite_relative_error} shows that the finite-flake result closely reproduces the infinite-lattice propagator throughout the interior, while the deviation increases as the drain approaches the boundary. The finite graphical recurrence therefore approaches the bulk infinite-lattice map when both contacts are sufficiently far from the finite edge.

\begin{figure}[t]
\centering
\includegraphics[width=0.98\columnwidth]{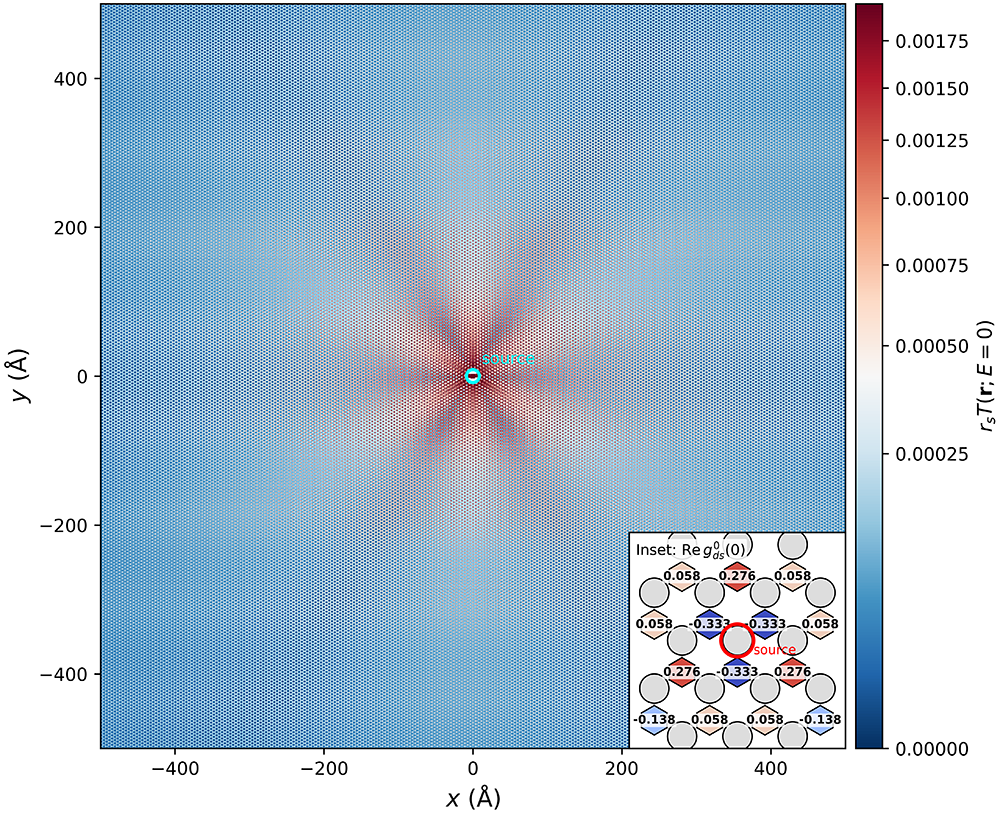}
\caption{Zero-energy transmission map of pristine infinite graphene for an $A$-sublattice source. The main illustration shows $r_sT_0(\mathbf r;E=0)$ for a Gaussian tip, and the inset shows the near-source reference amplitude $\mathrm{Re}\,g_{ds}^{0}(0)$.}
\label{fig:infinite_pristine_transmission}
\end{figure}

To represent a scanning probe with finite spatial resolution, we introduce a normalized tip orbital centered at the continuous position $\mathbf r$,
\begin{equation}
|\chi_{\mathbf r}\rangle
=
\sum_i w_i(\mathbf r)|i\rangle,
\quad
\sum_i|w_i(\mathbf r)|^2=1,
\end{equation}
with Gaussian weights
\begin{equation}
w_i(\mathbf r)
\propto
\exp\left[-\frac{|\mathbf r-\mathbf R_i|^2}{4\sigma_{\rm tip}^2}\right].
\label{eq:gaussian_tip_weights}
\end{equation}
The coherent drain amplitude is
\begin{equation}
G_{\chi s}(z;\mathbf r)
\equiv
\langle\chi_{\mathbf r}|G(z)|s\rangle
=
\sum_i w_i^*(\mathbf r)G_{is}(z),
\end{equation}
and the pristine scanning transmission is denoted by
\begin{equation}
T_0(\mathbf r;E)
\equiv
\Gamma_s(E)\Gamma_{\rm tip}(E)
|G_{\chi s}^{(0)}(E;\mathbf r)|^2.
\label{eq:pristine_scanning_transmission}
\end{equation}
Figure~\ref{fig:infinite_pristine_transmission} shows this map for $\sigma_{\rm tip}=1.4~\text{\AA}$. The plotted factor $r_s=|\mathbf r-\mathbf r_s|$, where $\mathbf r_s$ is the source position, compensates the leading spatial decay and exposes the far-field angular structure.

\subsection{Local defects}
A local defect modifies the infinite host only within a finite defect subspace. Let $D$ denote the defective sites and
\begin{equation}
V_D
=
\sum_{c,c'\in D}
|c\rangle(V_D)_{cc'}\langle c'|
\end{equation}
be the corresponding perturbation. Dyson's equation gives the finite-rank correction~\cite{Economou2006}
\begin{equation}
g_{D;ds}^\infty(z)
=
g_{0;ds}^\infty(z)
+
g_{0;dD}^\infty(z)
\mathcal T_D(z)
g_{0;Ds}^\infty(z),
\label{eq:infinite_defect_tmatrix_green}
\end{equation}
where
\begin{equation}
\mathcal T_D(z)
=
V_D\left[I-g_{0;DD}^\infty(z)V_D\right]^{-1}.
\label{eq:infinite_defect_tmatrix}
\end{equation}
All entries in Eq.~\eqref{eq:infinite_defect_tmatrix_green} are therefore built from the pristine reference map, and the only inversion is restricted to the finite defect set.

A vacancy at $v$ is obtained from an onsite potential $(V_D)_{vv}\rightarrow\infty$. For the single vacancy,
\begin{equation}
g_{v;ds}^\infty(z)
=
g_{0;ds}^\infty(z)
-
\frac{
g_{0;dv}^\infty(z)g_{0;vs}^\infty(z)}
{g_{0;vv}^\infty(z)}.
\label{eq:single_infinite_vacancy_green}
\end{equation}
Because $g_{0;vv}^\infty(0)=0$, the vacancy expression must be evaluated at finite $z=i\eta$ before taking the zero-energy limit. This singular behavior is the Green-function signature of the vacancy-induced zero-energy state associated with sublattice imbalance~\cite{Pereira2006,Pereira2008}.

\begin{figure}[t]
\centering
\includegraphics[width=0.98\columnwidth]{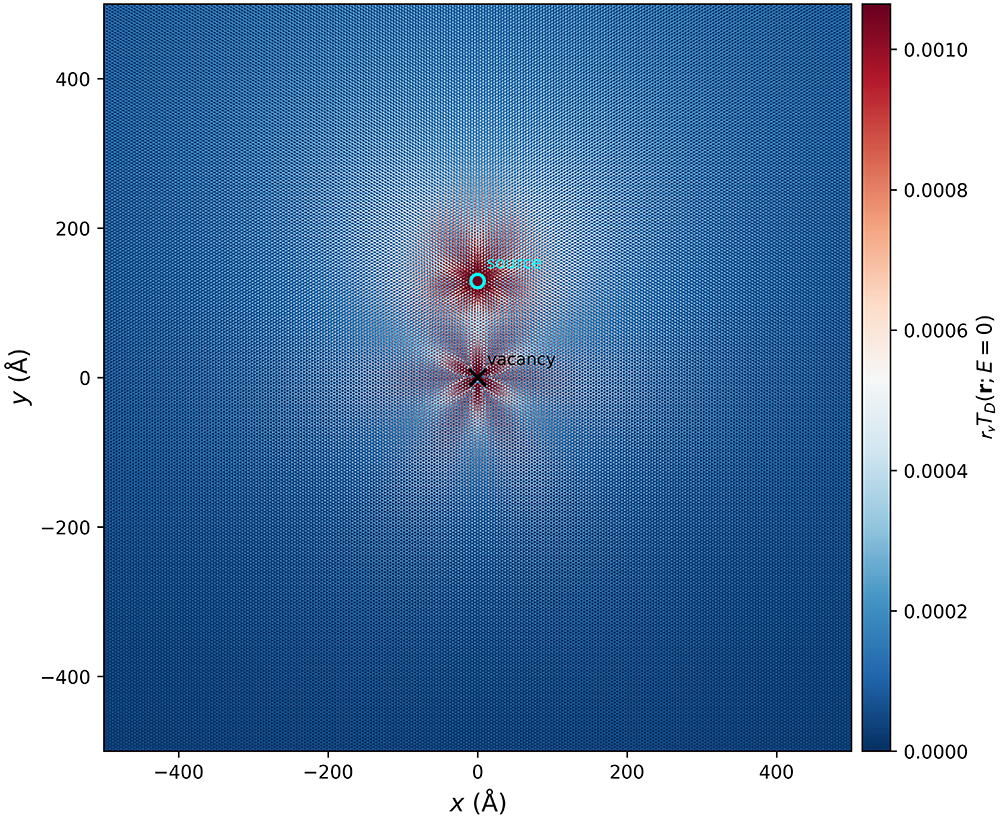}
\caption{Zero-energy transmission map of infinite graphene with a single $B$-sublattice vacancy. The illustration shows $r_vT_D(\mathbf r;E=0)$ for a Gaussian tip with $\sigma_{\rm tip}=1.4~\text{\AA}$, evaluated at $\eta=10^{-3}|t|$.}
\label{fig:infinite_vacancy_transmission}
\end{figure}

Using the same normalized Gaussian tip orbital, we denote the defected scanning transmission by
\begin{equation}
T_D(\mathbf r;E)
\equiv
\Gamma_s(E)\Gamma_{\rm tip}(E)
|G_{\chi s}^{(D)}(E;\mathbf r)|^2.
\end{equation}
Figure~\ref{fig:infinite_vacancy_transmission} shows the result for a single $B$-sublattice vacancy at position $\mathbf r_v$, with the $A$-sublattice source displaced along the armchair direction. The factor $r_v=|\mathbf r-\mathbf r_v|$ compensates the spatial decay relative to the vacancy, while the finite value $\eta=10^{-3}|t|$ regularizes the vacancy-induced pole.

The finite and infinite constructions therefore share the same graphical logic: we first generate a reference Green-function map and then incorporate the finite set of local degrees of freedom that distinguishes the target structure from that reference. This approach also connects naturally to double-tip scanning tunneling microscopy (STM). Niu et al. showed that, for noninteracting electrons at zero temperature, the inter-tip transconductance is proportional to the squared modulus of the sample's retarded Green function between the tip positions to leading order in the tip--sample tunneling couplings~\cite{Niu1995}. Fixing one tip and scanning the other thus provides a spatial probe of coherent propagation and defect-induced interference. The graphical rules developed here provide a framework for interpreting such spatial patterns at zero energy in terms of lattice connectivity, finite boundaries, and local defects.

\section{Discussion}
\label{sec:discussion_conclusion}

In the previous analysis, some steps such as the use of Yanghui-triangle coefficients, the three-neighbor recurrence, and the $C_3$ sector structure are specific to the honeycomb lattice. The broader construction, however, relies only on bipartite connectivity and a usable reference Green-function map. For another bipartite lattice, the local recurrence and the resulting reference pattern will change, but the subsequent steps remain the same, including local degrees of freedom generate spatial profiles through the reference map, a finite effective matrix selects the surviving zero modes, and zero-mode projectors correct the same- and opposite-sublattice channels.

The exact rules derived here assume nearest-neighbor intersublattice hopping, vanishing onsite energies, chiral symmetry, and the zero-energy limit. Same-sublattice hopping, finite onsite terms, or finite energy generally open the diagonal Green-function blocks and replace the strict local rule by an energy-dependent recurrence. The present construction should therefore be understood as the chiral zero-energy reference point from which such perturbations can be analyzed. The local-contact expressions also assume single-site source and drain self-energies; extended contacts can be included by replacing the scalar two-contact relation with its matrix counterpart.

The main value of the graphical formulation is not merely computational. It separates the roles of the boundary, the reference connectivity, and the zero modes. The boundary selects the directed sector of the reference pattern; the reference Green function carries amplitudes between local connection points; and the finite matrix $M$ records whether the profiles generated by the local sites can coexist as exact zero modes. This separation makes it possible to identify which structural feature opens, closes, or redirects a zero-energy transport channel without treating the full Hamiltonian as an opaque inverse.

The same structure also suggests a practical route to QI design. Once a reference map has been constructed, many local modifications can be screened by evaluating only the corresponding finite matrix $M$ and the resulting projector corrections. Candidate structures can therefore be compared in terms of the location and sublattice character of their surviving zero modes, rather than only through complete numerical transmission calculations.

\section{Conclusion}
\label{sec:conclusion}



We have formulated a graphical rule for constructing zero-energy QI patterns in bipartite lattices. For a unit-cell-replicated honeycomb flake with a fixed source, the finite-boundary conditions and the local recurrence jointly determine the zero-amplitude region and the directed propagation pattern. Within an untruncated part of the directed sector, the amplitudes follow the Yanghui-triangle rule. The corresponding isolated Green-function map determines the transmission map through the exact two-contact identity.

A general finite flake is decomposed into a unit-cell-replicated reference flake with an invertible hopping block and a finite set of local sites. The reference map extends amplitudes assigned to these local sites into spatial profiles, while the finite matrix $M$ selects the combinations that satisfy the zero-energy equations of the full flake. The projectors onto these full-flake zero modes determine the same-sublattice pole contributions and the contact conditions for allowed channels. When opposite-sublattice transport is allowed, the corresponding amplitudes are obtained from the source-driven recurrence with the local coupling conditions.

For an infinite lattice, we first construct the reference Green-function map of the lattice without defects using translational symmetry. This map gives the propagation amplitudes between lattice sites. Vacancies and other local defects are then incorporated through finite-rank corrections, requiring matrix inversion only within the finite defect subspace.

The resulting framework turns the zero-energy transport problem into a sequence of spatial operations that can be read directly from lattice connectivity. It therefore provides both a predictive construction of the QI map and a transparent interpretation of how boundaries, local modifications, and zero modes control coherent transport.

\appendix

\section{Invertibility of the unit-cell-replicated hopping block}
\label{app:h0_invertibility}

For the honeycomb connectivity used in Sec.~\ref{sec:graphical_rule}, the reference hopping block can be written as
\begin{equation}
h_0=-t(I+L_m+L_n).
\end{equation}
Ordering the unit cells by increasing $m+n$ makes both $L_m$ and $L_n$ strictly triangular in the same direction. Their sum
\begin{equation}
N=L_m+L_n
\end{equation}
is therefore strictly triangular and, in a finite-dimensional reference space, nilpotent. Thus $N^p=0$ for some positive integer $p$, and
\begin{equation}
h_0^{-1}
=-\frac{1}{t}
\sum_{\ell=0}^{p-1}(-N)^\ell.
\end{equation}
Hence $h_0$ is invertible for $t\neq0$.

\section{Zero-mode channel conditions}
\label{app:zero_mode_channels}

\subsection{Same-sublattice channel}

For two $A$-sublattice contacts, define
\begin{align}
a=\langle s_A|\Pi_0^A|s_A\rangle,
\notag\\
b=\langle d_A|\Pi_0^A|d_A\rangle,
\notag\\
c=\langle d_A|\Pi_0^A|s_A\rangle.
\end{align}
Near zero energy,
\begin{align}
g_{ss}=\frac{a}{z}+O(1),
\notag\\
g_{dd}=\frac{b}{z}+O(1),
\notag\\
g_{ds}=\frac{c}{z}+O(1).
\end{align}
The most singular term in the denominator of Eq.~\eqref{eq:two_contact_identity} is proportional to $(ab-|c|^2)/z^2$. Using $\mathbf p_s^A=\Phi_A^\dagger|s_A\rangle$ and $\mathbf p_d^A=\Phi_A^\dagger|d_A\rangle$, one has
\begin{align}
&a=(\mathbf p_s^A)^\dagger K^{-1}\mathbf p_s^{A},
\quad
b=(\mathbf p_d^A)^\dagger K^{-1}\mathbf p_d^{A},
\notag\\
&c=(\mathbf p_d^A)^\dagger K^{-1}\mathbf p_s^{A},
\quad
K = \Phi^\dagger\Phi.
\end{align}
The Cauchy--Schwarz inequality gives $ab-|c|^2\geq0$, with equality if and only if the two projection vectors are linearly dependent. A finite nonzero same-sublattice transmission therefore requires both vectors to be nonzero and linearly dependent, which gives Eq.~\eqref{eq:same_sublattice_channel_rule}.

\subsection{Opposite-sublattice channel and regular solution}

For an $A\rightarrow B$ channel, the off-diagonal isolated Green-function element is regular, whereas
\begin{align}
g_{ss}=\frac{\langle s_A|\Pi_0^A|s_A\rangle}{z}+O(1),
\notag\\
g_{dd}=\frac{\langle d_B|\Pi_0^B|d_B\rangle}{z}+O(1).
\end{align}
A nonzero pole at either contact makes the denominator of Eq.~\eqref{eq:two_contact_identity} divergent and suppresses the regular opposite-sublattice contribution. The necessary conditions are therefore $\Pi_0^A|s_A\rangle=0$ and $\Pi_0^B|d_B\rangle=0$. In particular, when the source overlaps the $A$-sublattice zero-mode space, every $A\rightarrow B$ channel is closed and no opposite-sublattice map is needed for the transmission problem.

For completeness, define $|F(z)\rangle=g_F(z)|s_A\rangle$. The resolvent equation
\begin{equation}
(zI-H_F)|F(z)\rangle=|s_A\rangle
\label{eq:full_source_resolvent_equation}
\end{equation}
and the decomposition in Eq.~\eqref{eq:zero_mode_green_decomposition} give, in the limit $z\rightarrow0$,
\begin{equation}
-h_F|F_B^{\rm reg}\rangle=(I_A-\Pi_0^A)|s_A\rangle.
\label{eq:projected_regular_equation}
\end{equation}
For a reference-flake amplitude, define $\mathcal A_{m,n}^{\rm reg}=\langle b_{m,n}|F_B^{\rm reg}\rangle$ and $\Pi_{m,n}^A=\langle a_{m,n}|\Pi_0^A|s_A\rangle$. Away from the local connection points, Eq.~\eqref{eq:projected_regular_equation} becomes
\begin{equation}
t\left(\mathcal A_{m,n}^{\rm reg}+\mathcal A_{m-1,n}^{\rm reg}+\mathcal A_{m,n-1}^{\rm reg}\right)=\delta_{m0}\delta_{n0}-\Pi_{m,n}^A.
\label{eq:zero_mode_corrected_recurrence}
\end{equation}
When the opposite-sublattice branch is allowed, $\Pi_0^A|s_A\rangle=0$, so $\Pi_{m,n}^A=0$ and Eq.~\eqref{eq:projected_regular_equation} reduces to the ordinary source equation
\begin{equation}
-h_F|F_B^{\rm reg}\rangle=|s_A\rangle.
\label{eq:ordinary_regular_source_equation}
\end{equation}
The local recurrence then has exactly the source-driven form of Eq.~\eqref{eq:local_recursion_minimal}, while $W_A$, $W_B$, and $X$ provide the corresponding conditions at the local connection points.

Because $h_F$ is singular, a $B$-sublattice zero mode may be added to any particular solution of Eq.~\eqref{eq:ordinary_regular_source_equation}. The regular Green function fixes this freedom through
\begin{equation}
\Pi_0^B|F_B^{\rm reg}\rangle=0.
\label{eq:regular_orthogonality_condition}
\end{equation}
This global choice is not required to determine the amplitudes at the allowed drains. If two particular solutions differ by a $B$-sublattice zero mode $|\chi_B\rangle$, then $\Pi_0^B|d_B\rangle=0$ implies $\langle d_B|\chi_B\rangle=0$. The amplitude at an allowed drain is therefore independent of the added zero-mode component and can be obtained from the ordinary source-driven recurrence. This is why the main-text graphical procedure needs only the source recurrence together with the exclusion of drains lying in the $B$-zero-mode scope.

\section{Closed-form infinite-graphene reference propagator}
\label{app:infinite_closed_forms}

Using the coordinates
\begin{equation}
q_1=m,
\quad
q_2=n,
\quad
q_3=-m-n-1,
\end{equation}
a cyclic permutation can place every lattice point into one of two inequivalent regions. In an acute sector, the coordinates can be written as
\begin{equation}
(q_1,q_2,q_3)=(a,b,-a-b-1),
\quad
a,b\geq0,
\end{equation}
and the zero-energy propagator is
\begin{equation}
\mathcal G_{\rm ac}(a,b)
=
\frac{(-1)^b}{\pi t}
[\mathcal G_{\rm ac,1}(a,b)
+
\mathcal G_{\rm ac,2}(a,b)],
\label{eq:appendix_acute_sector}
\end{equation}
where
\begin{align}
&\mathcal G_{\rm ac,1}(a,b)
=
(-1)^a
\binom{a+b}{b}
\frac{\pi}{3},
\notag\\
&\mathcal G_{\rm ac,2}(a,b)
=
\sum_{\substack{l=-b\\l\neq0}}^{a}
(-1)^{a-l}
\binom{a+b}{b+l}
\frac{\sin(l\pi/3)}{l}.
\end{align}

In the region between two neighboring acute sectors,
\begin{equation}
(q_1,q_2,q_3)=(-a-1,-b-1,a+b+1),
\quad
a,b\geq0,
\end{equation}
and
\begin{equation}
\mathcal G_{\rm int}(a,b)
=
\frac{(-1)^{b+1}}{\pi t}
\sum_{l=0}^{b}
\binom{b}{l}
\frac{\sin\theta_l}
{a+b+1-l},
\label{eq:appendix_between_sector}
\end{equation}
where
\begin{equation}
\theta_l
=
\frac{2\pi}{3}(a+b+1-l).
\end{equation}
Together with the cyclic relation in Eq.~\eqref{eq:C3_cyclic_green}, Eqs.~\eqref{eq:appendix_acute_sector} and \eqref{eq:appendix_between_sector} determine $G(m,n)$ throughout the infinite lattice.

\begin{acknowledgments}
This work is supported by the National Key R\&D Program of China (Grant No. 2023YFB4402600), and the Shanghai Technical Service Center of Science and Engineering Computing, Shanghai University. X.H.F. is also grateful to dear friend Jiangbo Zhao for generously sharing his time and for many thoughtful discussions.
\end{acknowledgments}

\bibliography{refs}

\end{document}